\documentclass[10pt,a4paper]{article}
\usepackage[utf8]{inputenc}
\usepackage[english]{babel}
\usepackage{amsmath}
\usepackage{amsfonts}
\usepackage{amssymb}

\usepackage{hyperref}
\usepackage[numbers]{natbib}
\usepackage{xcolor}
\allowdisplaybreaks

\begin{document}
\title{Commutation relations for functions of canonical conjugate operators}
\author{Conrado Badenas}
\maketitle

\section{Motivation}

Page 106 of \cite{Siegel:Fields4} (and page 3 of \cite{Siegel:Fields1}) includes the expression
\begin{equation}\label{siegelEq}
AB-BA=-i\hbar\sum_{m=1}^N\left(\frac{\partial A}{\partial p_m}\frac{\partial B}{\partial q^m}-\frac{\partial B}{\partial p_m}\frac{\partial A}{\partial q^m}\right)+\mathcal{O}(\hbar^2)
\end{equation}
for $A$ and $B$ a pair of functions of $N$ pairs of canonical conjugate operators $p_m$ and $q^m$. This expression is similar to the well-known procedure of canonical quantization (eq. 11 of \cite{Dirac:1925})
\begin{equation*}
AB-BA=i\hbar\{A,B\},
\end{equation*}
where
\begin{equation*}
\{A,B\}:=-\sum_{m=1}^N\left(\frac{\partial A}{\partial p_m}\frac{\partial B}{\partial q^m}-\frac{\partial B}{\partial p_m}\frac{\partial A}{\partial q^m}\right)
\end{equation*}
is Poisson bracket for two functions ($A$ and $B$) of $N$ pairs of canonical variables $p_m$ and $q^m$ on phase space.


In above expressions, symbol $A$ (also, $B$) is used ambiguosly for both an operator (in a Hilbert space) in quantum mechanics, and a (real-valued) function in classical mechanics. Moreover, $p_m$ (and $q^m$) has been described as an operator (with a non-commuting product) and a real number (with a commuting product). In the following, I will suppose that these symbols represent elements of a noncommutative associative algebra $\mathcal{A}$ over a field, thus special care on the order of products will be taken.

The motivation of this text is to understand better and prove \eqref{siegelEq}, which is not hard: in section 2, just reorder at convenience any term that is already multiplied by $\hbar$, and add $\mathcal{O}(\hbar^2)$ at the end of the expression if not already there. But this procedure does not compute the expression of $\mathcal{O}(\hbar^2)$ for \eqref{siegelEq}, which is unsatisfactory for those who want to know more. In order to achieve this, \cite{TranstrumVanhuele:2005} has been essential: their methods are applied in section 3 to obtain the infinite series of powers of $\hbar$ in \eqref{siegelEq}. In section 4, the class of functions $A,B$ is expanded to include negative integer powers of $p_m$ and $q^m$.

\section{The proof of equation \ref{siegelEq}}

Let $A,B,C,D$ be elements in a noncommutative associative algebra $\mathcal{A}$ over a field $\mathcal{F}$, and let $a,b,c,d$ be elements in $\mathcal{F}$. Among the properties of this algebra I highlight these:
\begin{itemize}
\item $aA,A+B,-A,AB\in\mathcal{A}$,
\item $A+B=B+A$,
\item in general $AB\neq BA$ (although $AB=BA$ for some $A,B$),
\item $0$ is a special element in $\mathcal{A}$ such that $A+0=A$ and $A0=0A=0$,
\item $0$ is a special element\footnote{Same symbol as before, but a different element. Distinction is possible by context.} in $\mathcal{F}$ such that $0A=0$ and $0a=0$,
\item $I$ is a special element in $\mathcal{A}$ such that $AI=IA=A$,
\item $1$ is a special element in $\mathcal{F}$ such that $1A=A$,
\item $A(B+C)=AB+AC$, and $(A+B)C=AC+BC$.
\end{itemize}
Then, commutator $[\cdot,\cdot]:\mathcal{A}\times\mathcal{A}\rightarrow\mathcal{A}$ is defined as $[A,B]:=AB-BA$, and these properties can easily be proven:
\begin{itemize}
\item $[aA+bB,cC+dD]=ac[A,C]+ad[A,D]+bc[B,C]+bd[B,D]$ (linearity),
\item $[A,B]=-[B,A]$ (anticommutativity),
\item $[A,[B,C]]+[B,[C,A]]+[C,[A,B]]=0$ (Jacobi identity),
\item $[A,BC]=[A,B]C+B[A,C]$ (Leibniz rule for $A$), and
\item $[A,I]=0$.
\end{itemize}
This property will be used extensively:
\begin{align*}
[AB,CD] &= [A,C]BD + C[A,D]B + A[B,C]D + AC[B,D] - [A,C][B,D] 
\\&= [A,C]DB + C[A,D]B + A[B,C]D + AC[B,D]
\\&= [A,C]BD + C[A,D]B + A[B,C]D + CA[B,D].
\end{align*}

Let $p,q$ be two elements in $\mathcal{A}$ such that $[p,q]=cI,c\in\mathcal{F}$.\footnote{In quantum mechanics, $\mathcal{F}=\mathbb{C}$ and $c=-i\hbar$ ($i^2=-1$ and $\hbar:=6.626\;070\;15\times10^{-34}/2\pi$ J/Hz) when $p$ stands for momentum operator and $q$ for position operator.} Now let $A^p(p) := \sum_i a_i^p p^i$ and $B^q(q) := \sum_j b_j^q q^j$ define operations with $p$ and $q$ respectively, where $a_i^p,b_j^q\in\mathcal{F}$. Note that $A^p(p),B^q(q)\in\mathcal{A}$ but $A^p,B^q$ are not functions from $\mathcal{A}$ to $\mathcal{A}$ because their domain is a one-element set ($A^p:\{p\}\longrightarrow\mathcal{A}$ and $B^q:\{q\}\longrightarrow\mathcal{A}$). Nevertheless, some concepts for functions will be used for $A^p,B^q$, such as $\partial_p\sum_i a_i^p p^i:=\sum_i a_i^p i p^{i-1}$. Now, $[A^p,B^q]$ can be written in terms of $[p,q]$, through some manipulations:\footnote{In previous and next expressions, $i,j,\ldots\in\mathbb{N}:=\{0,1,2,\ldots\}$ when $i,j,\ldots$ are used as counters or exponents/powers, or are related to exponents/powers.}
\begin{align*}
[p^2,q] &= [p,q]p+p[p,q] = cp+pc = c2p =: c\partial_p p^2 ;
\\
\text{if }[p^i,q] &= cip^{i-1} =: c\partial_p p^i \text{, then}
\\
[p^{i+1},q] &= [p^i,q]p+p^i[p,q] = cip^{i-1}p+p^ic
\\
&= c(i+1)p^i =: c\partial_p p^{i+1} \quad \forall i\in\mathbb{N} ;
\\
[A^p,q] &= \sum_i a_i^p[p^i,q] = c\sum_i a_i^p ip^{i-1} =: c\partial_p A^p ;
\end{align*}
\begin{align*}
[p,q^2] &= [p,q]q+q[p,q] = cq+qc = c2q =: c\partial_q q^2 ;
\\
\text{if }[p,q^j] &= cjq^{j-1} =: c\partial_q q^j \text{, then}
\\
[p,q^{j+1}] &= [p,q^j]q+q^j[p,q] = cjq^{j-1}q+q^jc
\\
&= c(j+1)q^j =: c\partial_q q^{j+1} \quad \forall j\in\mathbb{N} ;
\\
[p,B^q] &= \sum_j b_j^q[p,q^j]=c\sum_j b_j^q jq^{j-1}=:c\partial_q B^q ;
\end{align*}
\begin{align*}
[p^2,q^2] &= [p,q^2]p+p[p,q^2] = 2cqp+p2cq
\\
&= 2cpq-2c[p,q]+2cpq = 4cpq-2c^2I
\\
&= c(\partial_p p^2)(\partial_q q^2) + \mathcal{O}(c^2)\text{, where }\mathcal{O}(c^2) = -2c^2I;
\\
[p^2,q^j] &= [p,q^j]p+p[p,q^j] = c(\partial_q q^j)p + cp(\partial_q q^j)
\\
&= cp(\partial_q q^j) + \mathcal{O}(c^2) + cp(\partial_q q^j) = c(2p)(\partial_q q^j) + \mathcal{O}(c^2)
\\
&= c(\partial_p p^2)(\partial_q q^j) + \mathcal{O}(c^2) \quad \forall j\in\mathbb{N} ;
\\
\text{if }[p^i,q^j] &= c(\partial_p p^i)(\partial_q q^j) + \mathcal{O}(c^2)  \quad \forall j\in\mathbb{N}\text{, then}
\\
[p^{i+1},q^j] &= [p,q^j]p^i+p[p^i,q^j] = c(\partial_q q^j)p^i + cp(\partial_p p^i)(\partial_q q^j) + \mathcal{O}(c^2)
\\
&= cp^i(\partial_q q^j) + \mathcal{O}(c^2) + cip^i(\partial_q q^j) + \mathcal{O}(c^2)
\\
&= c(i+1)p^i(\partial_q q^j) + \mathcal{O}(c^2)
\\
&= c(\partial_p p^{i+1})(\partial_q q^j) + \mathcal{O}(c^2) \quad \forall i,j\in\mathbb{N} \text{; and}
\end{align*}
\begin{align}\label{ABplusOc2}
[A^p,B^q] &= \sum_{i,j} a_i^p b_j^q [p^i,q^j] = c\sum_{i,j} a_i^p b_j^q (\partial_p p^i)(\partial_q q^j) + \mathcal{O}(c^2)\notag
\\
&= c(\partial_p A^p)(\partial_q B^q) + \mathcal{O}(c^2) .
\end{align}

Next, let extend the concept of $A^p(p)$ and $B^q(q)$ to $A^q(q)$ and $B^p(p)$: Let $A^r(r):=\sum_i a_i^r r^i$, $B^r(r):=\sum_j b_j^r r^j$, where $r$ is a label that stands for $p$ or $q$ at convenience, and $a_i^r,b_j^r\in\mathcal{F}$. The goal is to compute $[A^p(p)A^q(q),B^p(p)B^q(q)]$ up to order $c$, by using \eqref{ABplusOc2}, reordering any term already multiplied by $c$, and throwing the trash in $\mathcal{O}(c^2)$. Note that $[A(r),B(r)]=0$ for every pair of functions\footnote{Remember that domain of these functions is a one-element set, and that a ``derivative'' such as $(\partial_p A^p q^i)$ is a convenient way to write $\sum_k (a_k^p k p^{k-1} q^i)$ because these larger expressions share some properties with actual derivatives.} $A,B$ of $r$, and every $r$ in $\mathcal{A}$:
\begin{align*}
[A^p q^i,B^q] &= [A^p,B^q]q^i + A^p\underbrace{[q^i,B^q]}_{=0} = c(\partial_p A^p)(\partial_q B^q)q^i + \mathcal{O}(c^2)
\\
&= c(\partial_p A^p)q^i(\partial_q B^q) + \mathcal{O}(c^2) = c(\partial_p A^p q^i)(\partial_q B^q) + \mathcal{O}(c^2) ;
\\
[A^p A^q,B^q] &= \sum_i a_i^q [A^p q^i,B^q] = \sum_i a_i^q c(\partial_p A^p q^i)(\partial_q B^q) + \mathcal{O}(c^2)
\\
&= c(\partial_p A^p A^q)(\partial_q B^q) + \mathcal{O}(c^2) ;
\\
[A^p,p^j B^q] &= \underbrace{[A^p,p^j]}_{=0} B^q + p^j[A^p,B^q] = cp^j(\partial_p A^p)(\partial_q B^q) + \mathcal{O}(c^2)
\\
&= c(\partial_p A^p)p^j(\partial_q B^q) + \mathcal{O}(c^2) = c(\partial_p A^p)(\partial_q p^j B^q) + \mathcal{O}(c^2) ;
\\
[A^p,B^p B^q] &= \sum_j b_j^p [A^p,p^j B^q] = \sum_j b_j^p c(\partial_p A^p)(\partial_q p^j B^q) + \mathcal{O}(c^2)
\\
&= c(\partial_p A^p)(\partial_q B^p B^q) + \mathcal{O}(c^2) \text{; and}
\\
[A^p A^q,B^p B^q] &= [A^p A^q,B^p]B^q+B^p[A^p A^q,B^q]
\\
&= B^p[A^p A^q,B^q] - [B^p,A^p A^q]B^q
\\
&= B^p c(\partial_p A^p A^q)(\partial_q B^q) - c(\partial_p B^p)(\partial_q A^p A^q) B^q + \mathcal{O}(c^2)
\\
&= c\big( (\partial_p A^p A^q)(\partial_q B^p B^q) - (\partial_p B^p B^q)(\partial_q A^p A^q)\big) + \mathcal{O}(c^2).
\end{align*}

Now, I claim that any function $A(p,q)$ can be written as a sum of functions of type $A^p(p)A^q(q)$ (it can be named ``$pq$-order''), i.e., $A(p,q)=\sum_i A_i^p(p) A_i^q(q)$ (also, $B(p,q)=\sum_j B_j^p(p) B_j^q(q)$). It seems counterintuitive since, e.g., $qqpp$ is written in a reverse order (``$qp$-order''), and $qpqp$ is not written in any order. Fortunately, any (sum of) non-$pq$-ordered monomial(s) can be written as a sum of $pq$-ordered monomials, which is the basis of my claim. For example,
\begin{align*}
pqpq &= p(pq-cI)q = ppqq-cpq,
\\
qpqp &= (pq-cI)(pq-cI) = pqpq-2cpq+c^2I = ppqq-3cpq+c^2I,
\\
qqpp &= q(pq-cI)p=qpqp-cqp = qpqp-c(pq-cI)
\\
&= ppqq-3cpq+c^2I -cpq+c^2I = ppqq-4cpq+2c^2I.
\end{align*}
Thus, if for some reason the commutator of function $A(p,q) = 3pqpq -2 qqpp$ with some function $B(p,q)$ had to be calculated, then $A$ should be rewritten as $3(ppqq-cpq)-2(ppqq-4cpq+2c^2I)=ppqq+5cpq-4c^2I$, i.e., $A(p,q)=\sum_{i=1}^3 A_i^p A_i^q$ with $(A_1^p,A_1^q,A_2^p,A_2^q,A_3^p,A_3^q) = (pp,qq,5cp,q,-4c^2I,I)$; and the same should be done for any $B(p,q)$. Moreover, this choice for $A_i^p,A_i^q$ is not unique: last choice was made by reading the $pq$-ordered expression of $A$, but uncountably many other choices are possible, e.g. $(2cI$, $-2cI$, $-\pi pp$, $-qq/\pi$, $\sqrt 5p$, $\sqrt 5cq)$.

Therefore, in general,
\begin{align*}
[A,B] &= \sum_{i,j} [A_i^p A_i^q,B_j^p B_j^q] 
\\
&= c\sum_{i,j} \left((\partial_p A_i^p A_i^q) (\partial_q B_j^p B_j^q) - (\partial_p B_j^p B_j^q) (\partial_q A_i^p A_i^q)\right) + \mathcal{O}(c^2)
\\
&= c \left( (\partial_p {\textstyle \sum_i} A_i^p A_i^q) (\partial_q {\textstyle \sum_j} B_j^p B_j^q)
- (\partial_p {\textstyle \sum_j} B_j^p B_j^q) (\partial_q {\textstyle \sum_i} A_i^p A_i^q)\right) + \mathcal{O}(c^2);
\end{align*}
thus,
\begin{equation}\label{ApqBpqwithOc2}
[A(p,q),B(p,q)] = c \big( (\partial_p A)(\partial_q B) - (\partial_p B)(\partial_q A) \big) + \mathcal{O}(c^2).
\end{equation}

Finally, if $A$ (and $B$) is any function of $2N$ elements
\begin{equation}\label{p1q1pNqN}
\begin{split}
p_1,q^1,\ldots,p_N,q^N\in\mathcal{A}
\\
[p_m,q^n] = c\delta_m^n I
\\
[p_m,p_n] = [q^m,q^n]=0
\end{split}
\end{equation}
then $A$ can be written as
\begin{equation}\label{Ap1q1pNqN}
A(p_1,q^1,\ldots,p_N,q^N) = \sum_i \prod_{m=1}^N A_i^m(p_m,q^m) = \sum_i A_i^1\cdots A_i^N,
\end{equation}
where every $A_i^m$ is written $pq$-ordered as before. For example, if $N=2$ and $A=p_2 p_1 q^1 q^2 p_1 p_2 q^2 q^1$, then $A$ can be written following \eqref{Ap1q1pNqN} as $A=p_1 q^1 p_1 q^1 p_2 q^2 p_2 q^2$ $ = (p_1 p_1 q^1 q^1 -c p_1 q^1) (p_2 p_2 q^2 q^2 -c p_2 q^2)$ $ = \underbrace{p_1p_1q^1q^1p_2p_2q^2q^2}_{i=1}$ $ + \underbrace{-cp_1p_1q^1q^1p_2q^2}_{i=2}$ $ + \underbrace{-cp_1q^1p_2p_2q^2q^2}_{i=3}$ $ + \underbrace{c^2p_1q^1p_2q^2}_{i=4}$, and this choice can be made (among uncountably many) for the $A_i^m$ in \eqref{Ap1q1pNqN}: $(A_1^1 , A_1^2 , A_2^1 , A_2^2 , A_3^1 , A_3^2 , A_4^1 , A_4^2)$ $ = (p_1p_1q^1q^1$, $p_2p_2q^2q^2$, $p_1p_1q^1q^1$, $-cp_2q^2$, $-cp_1q^1$, $p_2p_2q^2q^2$, $-cp_1q^1$, $-cp_2q^2)$.\footnote{This choice is specially interesting because $A_1^1 = A_2^1 \sim A_1^2 = A_3^2$ and $A_3^1 = A_4^1 \sim A_2^2 = A_4^2$.}

Note that \eqref{ApqBpqwithOc2} is the case $N=1$ for \eqref{siegelEq}, so the base case is already proved. Next step is to suppose \eqref{siegelEq} is true for $N$ and prove it true for $N+1$ (which will be rewritten as $M$ to save space): Let $A(p_1,q^1,\ldots,p_{M},q^{M})=\sum_i A_i^1\cdots A_i^{M}$ and $B(p_1,q^1,\ldots,p_{M},q^{M})=\sum_j B_j^1\cdots B_j^{M}$. Then,
\begin{align*}
[A,B] &= \sum_{i,j} [A_i^{1}\cdots A_i^{N} A_i^{M},B_j^{1}\cdots B_j^{N} B_j^{M}]
\\
&= \sum_{i,j} \Big(
\underbrace{[A_i^{1}\cdots A_i^{N},B_j^{1}\cdots B_j^{N}]}_{\sim\mathcal{O}(c)}A_i^{M}B_j^{M} + B_j^{1}\cdots B_j^{N}\underbrace{[A_i^{1}\cdots A_i^{N},B_j^{M}]}_{=0}A_i^{M}
\\
&\qquad\quad+ A_i^{1}\cdots A_i^{N}\underbrace{[A_i^{M},B_j^{1}\cdots B_j^{N}]}_{=0}B_j^{M} + A_i^{1}\cdots A_i^{N} B_j^{1}\cdots B_j^{N}\underbrace{[A_i^{M},B_j^{M}]}_{\sim\mathcal{O}(c)}
\\
&\qquad\quad- \underbrace{[A_i^{1}\cdots A_i^{N},B_j^{1}\cdots B_j^{N}][A_i^{M},B_j^{M}]}_{\sim\mathcal{O}(c^2)} \Big)
\\
&= c \sum_{i,j} \sum_{m=1}^N \big( (\partial_{p_m} A_i^{1}\cdots A_i^{N}) (\partial_{q^m} B_j^{1}\cdots B_j^{N})
\\
&\qquad\qquad\quad- (\partial_{p_m} B_j^{1}\cdots B_j^{N}) (\partial_{q^m} A_i^{1}\cdots A_i^{N}) \big) A_i^{M}B_j^{M} + \mathcal{O}(c^2)
\\
&\quad+ c \sum_{i,j} A_i^{1}\cdots A_i^{N} B_j^{1}\cdots B_j^{N} \left( (\partial_{p_{M}} A_i^{M}) (\partial_{q^{M}} B_j^{M}) - (\partial_{p_{M}} B_j^{M}) (\partial_{q^{M}} A_i^{M}) \right)
\\
&\qquad\quad+ \mathcal{O}(c^2) + \mathcal{O}(c^2)
\\
&= c \sum_{i,j} \sum_{m=1}^N \big( (\partial_{p_m} A_i^{1}\cdots A_i^{M}) (\partial_{q^m} B_j^{1}\cdots B_j^{M}) 
\\
&\qquad\qquad\quad- (\partial_{p_m} B_j^{1}\cdots B_j^{M}) (\partial_{q^m} A_i^{1}\cdots A_i^{M}) \big)
\\
&\quad+ c \sum_{i,j} \big( (\partial_{p_{M}} A_i^{1}\cdots A_i^{M}) (\partial_{q^{M}} B_j^{1}\cdots B_j^{M}) 
\\
&\qquad\qquad- (\partial_{p_{M}} B_j^{1}\cdots B_j^{M}) (\partial_{q^{M}} A_i^{1}\cdots A_i^{M}) \big) + \mathcal{O}(c^2)
\\
&= c \sum_{m=1}^N \big( (\partial_{p_m} A) (\partial_{q^m} B) - (\partial_{p_m} B) (\partial_{q^m} A) \big)
\\
&\qquad+ c \big( (\partial_{p_{M}} A) (\partial_{q^{M}} B) - (\partial_{p_{M}} B) (\partial_{q^{M}} A) \big) + \mathcal{O}(c^2)
\\
&= c \sum_{m=1}^{M} \big( (\partial_{p_m} A) (\partial_{q^m} B) - (\partial_{p_m} B) (\partial_{q^m} A) \big) + \mathcal{O}(c^2),
\end{align*}
i.e., it is true for $N+1$. Therefore, \eqref{siegelEq} is true $\forall N\in\mathbb{N}$.

\section{Beyond equation \ref{siegelEq}}

Now, the procedures of \cite{TranstrumVanhuele:2005} are applied to obtain the complete version of \eqref{ABplusOc2}:\footnote{In next expressions, $\partial^k:=\underbrace{\partial\cdots\partial}_{k\text{ times}}$ and $\partial^0 A:=A\;\forall A\in\mathcal{A}$, as usual for actual derivatives.}
\begin{align*}
[A^p,q^2] &= [A^p,q]q+q[A^p,q] = c(\partial_p A^p)q + cq(\partial_p A^p)
\\
&= c(\partial_p A^p)q + c(\partial_p A^p)q - c[\partial_p A^p,q] = 2c(\partial_p A^p)q - c^2\partial_p \partial_p A^p \\
&= -\sum_{k=1}^2 \binom{2}{k}(-c)^k(\partial_p^k A^p)q^{2-k} ;
\\
\text{if }[A^p,q^j] &= -\sum_{k=1}^j \binom{j}{k}(-c)^k(\partial_p^k A^p)q^{j-k} \text{, then}
\\
[A^p,q^{j+1}] &= [A^p,q]q^j+q[A^p,q^j] = c(\partial_p A^p)q^j -\sum_{k=1}^j \binom{j}{k}(-c)^k q(\partial_p^k A^p)q^{j-k}
\\
&= c(\partial_p A^p)q^j -\sum_{k=1}^j \binom{j}{k}(-c)^k \left((\partial_p^k A^p)q-[\partial_p^k A^p,q]\right)q^{j-k}
\\
&= c(\partial_p A^p)q^j -\sum_{k=1}^j \binom{j}{k}(-c)^k \left((\partial_p^k A^p)q-c\partial_p^{k+1}A^p\right)q^{j-k}
\\
&= c(\partial_p A^p)q^j -\sum_{k=1}^j \binom{j}{k}(-c)^k (\partial_p^k A^p)q^{j+1-k}
\\
&\quad- \sum_{k=1}^j \binom{j}{k}(-c)^{k+1} (\partial_p^{k+1} A^p)q^{j-k}
\\
&= c(\partial_p A^p)q^j -\sum_{k=1}^j \binom{j}{k}(-c)^k (\partial_p^k A^p)q^{j+1-k}
\\
&\quad- \sum_{k=2}^{j+1} \binom{j}{k-1}(-c)^k (\partial_p^k A^p)q^{j+1-k}
\\
&= c(\partial_p A^p)q^j -j(-c)(\partial_p A^p)q^j -\sum_{k=2}^j \binom{j}{k}(-c)^k (\partial_p^k A^p)q^{j+1-k} 
\\
&\quad- \sum_{k=2}^j \binom{j}{k-1}(-c)^k (\partial_p^k A^p)q^{j+1-k} -(-c)^{j+1} (\partial_p^{j+1} A^p)
\\
&= -(j+1)(-c)(\partial_p A^p)q^j -\sum_{k=2}^j \left[\binom{j}{k}+\binom{j}{k-1}\right](-c)^k (\partial_p^k A^p)q^{j+1-k}
\\
&\quad- (-c)^{j+1} (\partial_p^{j+1} A^p)
\\
&= -(j+1)(-c)(\partial_p A^p)q^j -\sum_{k=2}^j \binom{j+1}{k}(-c)^k (\partial_p^k A^p)q^{j+1-k}
\\
&\quad- (-c)^{j+1} (\partial_p^{j+1} A^p)
\\
&= -\sum_{k=1}^{j+1} \binom{j+1}{k}(-c)^k (\partial_p^k A^p)q^{j+1-k} \quad \forall j\in\mathbb{N}.
\intertext{Moreover,}
[A^p,q^j] &= -\sum_{k=1}^j \frac{j!}{k!(j-k)!}(-c)^k(\partial_p^k A^p)q^{j-k}
\\
&= -\sum_{k=1}^j \frac{j\cdots(j-k+1)}{k!}(-c)^k(\partial_p^k A^p)q^{j-k}
\\
&= -\sum_{k=1}^j \frac{(-c)^k}{k!}(\partial_p^k A^p)(\partial_q^k q^j) \quad(=0\text{ for }j=0)\text{; so}
\\
[A^p,B^q] &= \sum_{j=0}^\infty b_j [A^p,q^j] = -\sum_{j=1}^\infty b_j \sum_{k=1}^j \frac{(-c)^k}{k!}(\partial_p^k A^p)(\partial_q^k q^j)
\\
&= -\sum_{j,k=1}^\infty \frac{(-c)^k}{k!}(\partial_p^k A^p) (\partial_q^k b_j q^j) = -\sum_{k=1}^\infty \frac{(-c)^k}{k!}(\partial_p^k A^p) (\partial_q^k \sum_{j=0}^\infty b_j q^j),
\end{align*}
where the sum for $k$ has been extended to $\infty$ because $\partial_q^k q^j = 0$ for $k>j$, and the last sum for $j$ has been extended to $0$ because $\partial_q^k q^0 = 0$ for $k>0$. Therefore, the complete version of \eqref{ABplusOc2} is
\begin{align}\label{ABwithoutOc2}
[A^p,B^q] &= -\sum_{k=1}^\infty \frac{(-c)^k}{k!}(\partial_p^k A^p)(\partial_q^k B^q)
\\
&= c(\partial_p A^p)(\partial_q B^q) -\frac{c^2}{2}(\partial_p^2 A^p)(\partial_q^2 B^q) + \cdots.\notag
\end{align}

Next, let $A(p,q):=\sum_i A_i^p(p)A_i^q(q)$ and $B(p,q):=\sum_j B_j^p(p)B_j^q(q)$, as in section 2. Now, considering that every function of $p$ commutes with any function of $p$, that every function of $q$ commutes with any function of $q$ (for example, $[A_i^p,B_j^p]=[A_i^q,B_j^q]=0$), and last equation, one obtains:
\begin{align*}
[A,B] &= \sum_{i,j} [A_i^p A_i^q,B_j^p B_j^q] = \sum_{i,j} \left( [A_i^p,B_j^p B_j^q] A_i^q + A_i^p [A_i^q,B_j^p B_j^q] \right)
\\
&= \sum_{i,j} \left( B_j^p [A_i^p,B_j^q] A_i^q - A_i^p [B_j^p,A_i^q] B_j^q \right)
\\
&= -\sum_{i,j} \sum_{k=1}^\infty \frac{(-c)^k}{k!} \left( B_j^p (\partial_p^k A_i^p)(\partial_q^k B_j^q) A_i^q - A_i^p (\partial_p^k B_j^p)(\partial_q^k A_i^q)  B_j^q \right)
\\
&= -\sum_{i,j} \sum_{k=1}^\infty \frac{(-c)^k}{k!} \left( (\partial_p^k A_i^p) B_j^p A_i^q (\partial_q^k B_j^q) - (\partial_p^k B_j^p) A_i^p B_j^q (\partial_q^k A_i^q) \right)
\\
&= -\sum_{i,j} \sum_{k=1}^\infty \frac{(-c)^k}{k!}  (\partial_p^k A_i^p) \left(A_i^q B_j^p + [B_j^p,A_i^q]\right) (\partial_q^k B_j^q)
\\
&\quad+ \sum_{i,j} \sum_{k=1}^\infty \frac{(-c)^k}{k!} (\partial_p^k B_j^p) \left(B_j^q A_i^p + [A_i^p,B_j^q]\right) (\partial_q^k A_i^q)
\\
&= -\sum_{i,j} \sum_{k=1}^\infty \frac{(-c)^k}{k!}  \left( (\partial_p^k A_i^p A_i^q) (\partial_q^k B_j^p B_j^q) + (\partial_p^k A_i^p) [B_j^p,A_i^q] (\partial_q^k B_j^q) \right)
\\
&\quad+ \sum_{i,j} \sum_{k=1}^\infty \frac{(-c)^k}{k!} \left((\partial_p^k B_j^p B_j^q) (\partial_q^k A_i^p A_i^q) + (\partial_p^k B_j^p) [A_i^p,B_j^q] (\partial_q^k A_i^q)\right)
\\
&= -\sum_{k=1}^\infty \frac{(-c)^k}{k!}  \left( (\partial_p^k A) (\partial_q^k B) + \sum_{i,j} \sum_{l=1}^\infty \frac{(-c)^l}{l!} (\partial_p^k A_i^p) (\partial_p^l B_j^p)(\partial_q^l A_i^q) (\partial_q^k B_j^q) \right)
\\
&\quad+ \sum_{k=1}^\infty \frac{(-c)^k}{k!} \left( (\partial_p^k B) (\partial_q^k A) + \sum_{i,j} \sum_{l=1}^\infty \frac{(-c)^l}{l!} (\partial_p^k B_j^p) (\partial_p^l A_i^p)(\partial_q^l B_j^q) (\partial_q^k A_i^q)\right)
\\
&= -\sum_{k=1}^\infty \frac{(-c)^k}{k!}  \left( (\partial_p^k A) (\partial_q^k B) - (\partial_p^k B) (\partial_q^k A) \right)
\\
&\quad- \sum_{i,j} \sum_{k,l=1}^\infty \frac{(-c)^k}{k!} \frac{(-c)^l}{l!} (\partial_p^k A_i^p) (\partial_p^l B_j^p)(\partial_q^l A_i^q) (\partial_q^k B_j^q)
\\
&\quad+ \sum_{i,j} \sum_{l,k=1}^\infty \frac{(-c)^l}{l!} \frac{(-c)^k}{k!} \underbrace{(\partial_p^k B_j^p) (\partial_p^l A_i^p)}_{=(\partial_p^l A_i^p)(\partial_p^k B_j^p)} \underbrace{(\partial_q^l B_j^q) (\partial_q^k A_i^q)}_{=(\partial_q^k A_i^q)(\partial_q^l B_j^q)}.
\end{align*}
Therefore, the complete version of \eqref{ApqBpqwithOc2} is
\begin{equation}\label{ApqBpqwithoutOc2}
[A(p,q),B(p,q)] = -\sum_{k=1}^\infty \frac{(-c)^k}{k!}  \left( (\partial_p^k A) (\partial_q^k B) - (\partial_p^k B) (\partial_q^k A) \right).
\end{equation}

Now, the goal is to obtain the complete version of \eqref{siegelEq}. First step is to propose a generalization of \eqref{ApqBpqwithoutOc2} (the case $N=1$) to any $N$. Thus, considering $A=A(p_1,q^1)$ and $B=B(p_1,q^1)$, last equation can be written as
\begin{equation*}
0 = -\sum_{k_1=0}^\infty \frac{(-c)^{k_1}}{{k_1}!}  \left( (\partial_{p_1}^{k_1} A) (\partial_{q^1}^{k_1} B) - (\partial_{p_1}^{k_1} B) (\partial_{q^1}^{k_1} A) \right),
\end{equation*}
which gives an insight on the general case. Then, let $A$ (and $B$) be a function of $2N$ elements as in \eqref{p1q1pNqN}, and write $A$ (and $B$) as in \eqref{Ap1q1pNqN}. The proposal for the complete version of \eqref{siegelEq} is:
\begin{equation*}
0 = -\sum_{\substack{k_1=0\\\ldots\\k_N=0}}^\infty \prod_{m=1}^N \frac{(-c)^{k_m}}{{k_m}!}  \left( (\partial_{p_1}^{k_1} \cdots \partial_{p_N}^{k_N} A) (\partial_{q^1}^{k_1} \cdots \partial_{q^N}^{k_N} B) - (\partial_{p_1}^{k_1} \cdots \partial_{p_N}^{k_N} B) (\partial_{q^1}^{k_1} \cdots \partial_{q^N}^{k_N} A) \right),
\end{equation*}
which can be rewriten more compactly as
\begin{equation}\label{fundamentalEq}
0 = -\sum_{\substack{k_{1\cdot N}\\=0}}^\infty \prod_{m=1}^N \frac{(-c)^{k_m}}{{k_m}!}  \left( (\partial_{p_{1\cdot N}}^{k_{1\cdot N}} A) (\partial_{q^{1\cdot N}}^{k_{1\cdot N}} B) - (\partial_{p_{1\cdot N}}^{k_{1\cdot N}} B) (\partial_{q^{1\cdot N}}^{k_{1\cdot N}} A) \right)
\end{equation}
with the new notation $k_{1\cdot N} \equiv k_1,\ldots,k_N$, $\partial_{p_{m\cdot n}}^{k_{m\cdot n}}\equiv\partial_{p_m}^{k_m} \cdots \partial_{p_n}^{k_n}$, and $\partial_{q^{m\cdot n}}^{k_{m\cdot n}}\equiv\partial_{q^m}^{k_m} \cdots \partial_{q^n}^{k_n}$, or
\begin{equation}\label{siegelEqwihoutOc2}
[A,B] = -\sum_{\substack{k_{1\cdot N}\\=0*}}^\infty \prod_{m=1}^N \frac{(-c)^{k_m}}{{k_m}!}  \left( (\partial_{p_{1\cdot N}}^{k_{1\cdot N}} A) (\partial_{q^{1\cdot N}}^{k_{1\cdot N}} B) - (\partial_{p_{1\cdot N}}^{k_{1\cdot N}} B) (\partial_{q^{1\cdot N}}^{k_{1\cdot N}} A) \right) ,
\end{equation}
where ``0*'' means ``avoiding the case $k_1=\cdots=k_N=0$'' (because such case is the term $[A,B]$ that has been placed at the left-hand side of the equation).

To prove \eqref{siegelEqwihoutOc2} for all $N$ by induction, first consider the base case proved, which is \eqref{ApqBpqwithoutOc2}. Then, suppose \eqref{siegelEqwihoutOc2} true for $N$ and compute $[A,B]$ for $N+1$ (which will be rewritten as $M$ to save space): Let $A(p_1,q^1,\ldots,p_{M},q^{M})=\sum_i A_i^1\cdots A_i^{M}\equiv\sum_i A_i^{1\cdot M}$ and $B(p_1,q^1,\ldots,p_{M},q^{M})=\sum_j B_j^1\cdots B_j^{M}\equiv\sum_j B_j^{1\cdot M}$, where $A_i^m$ and $B_j^m$ are functions of $p_m,q^m$. Then,
\begin{align*}
[A,B] &= \sum_{i,j} [A_i^{1\cdot N} A_i^{M},B_j^{1\cdot N} B_j^{M}]
\\
&= \sum_{i,j} \Big(
[A_i^{1\cdot N},B_j^{1\cdot N}] A_i^{M}B_j^{M} + B_j^{1\cdot N}\underbrace{[A_i^{1\cdot N},B_j^{M}]}_{=0} A_i^{M}
\\
&\qquad\quad+ A_i^{1\cdot N}\underbrace{[A_i^{M},B_j^{1\cdot N}]}_{=0} B_j^{M} + B_j^{1\cdot N} A_i^{1\cdot N} [A_i^{M},B_j^{M}] \Big)
\\
&=- \sum_{i,j} \sum_{\substack{k_{1\cdot N}\\=0*}}^\infty \prod_{m=1}^N \frac{(-c)^{k_m}}{{k_m}!}  \Big( (\partial_{p_{1\cdot N}}^{k_{1\cdot N}} A_i^{1\cdot N}) (\partial_{q^{1\cdot N}}^{k_{1\cdot N}} B_j^{1\cdot N})
\\
&\qquad\qquad\qquad\qquad\qquad\quad- (\partial_{p_{1\cdot N}}^{k_{1\cdot N}} B_j^{1\cdot N}) (\partial_{q^{1\cdot N}}^{k_{1\cdot N}} A_i^{1\cdot N}) \Big) A_i^{M}B_j^{M}
\\
&\quad- \sum_{i,j} \sum_{l=1}^\infty \frac{(-c)^l}{l!} B_j^{1\cdot N} A_i^{1\cdot N} \left( (\partial_{p_{M}}^l A_i^{M}) (\partial_{q^{M}}^l B_j^{M}) - (\partial_{p_{M}}^l B_j^{M}) (\partial_{q^{M}}^l A_i^{M}) \right)
\\
&= -\sum_{i,j} \sum_{\substack{k_{1\cdot N}\\=0*}}^\infty \prod_{m=1}^N \frac{(-c)^{k_m}}{{k_m}!}  (\partial_{p_{1\cdot N}}^{k_{1\cdot N}} A_i^{1\cdot N} A_i^{M}) (\partial_{q^{1\cdot N}}^{k_{1\cdot N}} B_j^{1\cdot N} B_j^{M})
\\
&\quad+ \sum_{i,j} \sum_{\substack{k_{1\cdot N}\\=0*}}^\infty \prod_{m=1}^N \frac{(-c)^{k_m}}{{k_m}!}  (\partial_{p_{1\cdot N}}^{k_{1\cdot N}} B_j^{1\cdot N}) (\partial_{q^{1\cdot N}}^{k_{1\cdot N}} A_i^{1\cdot N}) \left(B_j^{M}A_i^{M} - [B_j^{M},A_i^{M}] \right)
\\
&\quad- \sum_{i,j} \sum_{l=1}^\infty \frac{(-c)^l}{l!} \left( A_i^{1\cdot N} B_j^{1\cdot N} - [A_i^{1\cdot N},B_j^{1\cdot N}] \right) (\partial_{p_{M}}^l A_i^{M}) (\partial_{q^{M}}^l B_j^{M})
\\
&\quad+ \sum_{i,j} \sum_{l=1}^\infty \frac{(-c)^l}{l!} (\partial_{p_{M}}^l B_j^{1\cdot N} B_j^{M}) (\partial_{q^{M}}^l A_i^{1\cdot N} A_i^{M})
\\
&= -\sum_{i,j} \sum_{\substack{k_{1\cdot N}\\=0*}}^\infty \prod_{m=1}^N \frac{(-c)^{k_m}}{{k_m}!}  \Big( (\partial_{p_{1\cdot N}}^{k_{1\cdot N}} A_i^{1\cdot N} A_i^{M}) (\partial_{q^{1\cdot N}}^{k_{1\cdot N}} B_j^{1\cdot N} B_j^{M})
\\
&\qquad\qquad\qquad\qquad\qquad\quad- (\partial_{p_{1\cdot N}}^{k_{1\cdot N}} B_j^{1\cdot N} B_j^{M}) (\partial_{q^{1\cdot N}}^{k_{1\cdot N}} A_i^{1\cdot N} A_i^{M}) \Big)
\\
&\quad+ \sum_{i,j} \sum_{\substack{k_{1\cdot N}\\=0*}}^\infty \prod_{m=1}^N \frac{(-c)^{k_m}}{{k_m}!} (\partial_{p_{1\cdot N}}^{k_{1\cdot N}} B_j^{1\cdot N}) (\partial_{q^{1\cdot N}}^{k_{1\cdot N}} A_i^{1\cdot N})
\\
&\qquad\qquad\cdot \sum_{l=1}^\infty \frac{(-c)^l}{l!} \left( (\partial_{p_{M}}^l B_j^{M}) (\partial_{q^{M}}^l A_i^{M}) - (\partial_{p_{M}}^l A_i^{M}) (\partial_{q^{M}}^l B_j^{M}) \right)
\\
&\quad- \sum_{i,j} \sum_{l=1}^\infty \frac{(-c)^l}{l!} \Big( (\partial_{p_{M}}^l A_i^{1\cdot N} A_i^{M}) (\partial_{q^{M}}^l B_j^{1\cdot N} B_j^{M})
\\
&\qquad\qquad\qquad\qquad- (\partial_{p_{M}}^l B_j^{1\cdot N} B_j^{M}) (\partial_{q^{M}}^l A_i^{1\cdot N} A_i^{M}) \Big)
\\
&\quad- \sum_{i,j} \sum_{l=1}^\infty \frac{(-c)^l}{l!} \sum_{\substack{k_{1\cdot N}\\=0*}}^\infty \prod_{m=1}^N \frac{(-c)^{k_m}}{{k_m}!} \Big( (\partial_{p_{1\cdot N}}^{k_{1\cdot N}} A_i^{1\cdot N}) (\partial_{q^{1\cdot N}}^{k_{1\cdot N}} B_j^{1\cdot N})
\\
&\qquad\qquad\qquad\qquad- (\partial_{p_{1\cdot N}}^{k_{1\cdot N}} B_j^{1\cdot N}) (\partial_{q^{1\cdot N}}^{k_{1\cdot N}} A_i^{1\cdot N}) \Big) (\partial_{p_{M}}^l A_i^{M}) (\partial_{q^{M}}^l B_j^{M})
\\
&= -\sum_{\substack{k_{1\cdot N}\\=0*}}^\infty \prod_{m=1}^N \frac{(-c)^{k_m}}{{k_m}!}  \left( (\partial_{p_{1\cdot N}}^{k_{1\cdot N}} A) (\partial_{q^{1\cdot N}}^{k_{1\cdot N}} B) - (\partial_{p_{1\cdot N}}^{k_{1\cdot N}} B) (\partial_{q^{1\cdot N}}^{k_{1\cdot N}} A) \right)
\\
&\quad+ \sum_{i,j} \sum_{\substack{k_{1\cdot N}\\=0*}}^\infty \prod_{m=1}^N \frac{(-c)^{k_m}}{{k_m}!} \sum_{l=1}^\infty \frac{(-c)^l}{l!} (\partial_{p_{1\cdot N}}^{k_{1\cdot N}} B_j^{1\cdot N}) (\partial_{q^{1\cdot N}}^{k_{1\cdot N}} A_i^{1\cdot N}) (\partial_{p_{M}}^l B_j^{M}) (\partial_{q^{M}}^l A_i^{M})
\\
&\quad- \sum_{i,j} \sum_{\substack{k_{1\cdot N}\\=0*}}^\infty \prod_{m=1}^N \frac{(-c)^{k_m}}{{k_m}!} \sum_{l=1}^\infty \frac{(-c)^l}{l!} \underbrace{ (\partial_{p_{1\cdot N}}^{k_{1\cdot N}} B_j^{1\cdot N}) (\partial_{q^{1\cdot N}}^{k_{1\cdot N}} A_i^{1\cdot N}) (\partial_{p_{M}}^l A_i^{M}) (\partial_{q^{M}}^l B_j^{M}) }_\text{this term is cancelled...}
\\
&\quad- \sum_{l=1}^\infty \frac{(-c)^l}{l!} \left( (\partial_{p_{M}}^l A) (\partial_{q^{M}}^l B) - (\partial_{p_{M}}^l B) (\partial_{q^{M}}^l A) \right)
\\
&\quad- \sum_{i,j} \sum_{\substack{k_{1\cdot N}\\=0*}}^\infty \prod_{m=1}^N \frac{(-c)^{k_m}}{{k_m}!} \sum_{l=1}^\infty \frac{(-c)^l}{l!} (\partial_{p_{1\cdot N}}^{k_{1\cdot N}} A_i^{1\cdot N}) (\partial_{q^{1\cdot N}}^{k_{1\cdot N}} B_j^{1\cdot N}) (\partial_{p_{M}}^l A_i^{M}) (\partial_{q^{M}}^l B_j^{M})
\\
&\quad+ \sum_{i,j} \sum_{\substack{k_{1\cdot N}\\=0*}}^\infty \prod_{m=1}^N \frac{(-c)^{k_m}}{{k_m}!} \sum_{l=1}^\infty \frac{(-c)^l}{l!} \underbrace{ (\partial_{p_{1\cdot N}}^{k_{1\cdot N}} B_j^{1\cdot N}) (\partial_{q^{1\cdot N}}^{k_{1\cdot N}} A_i^{1\cdot N}) (\partial_{p_{M}}^l A_i^{M}) (\partial_{q^{M}}^l B_j^{M}) }_\text{...with this term}
\\
&= -\sum_{\substack{k_{1\cdot N}\\=0*}}^\infty \prod_{m=1}^N \frac{(-c)^{k_m}}{{k_m}!} \sum_{l=0}^0  \frac{(-c)^l}{l!} \Big( (\partial_{p_{1\cdot N}}^{k_{1\cdot N}} \partial_{p_{M}}^l A) (\partial_{q^{1\cdot N}}^{k_{1\cdot N}} \partial_{p_{M}}^l B)
\\
&\qquad\qquad\qquad\qquad\qquad\qquad\qquad- (\partial_{p_{1\cdot N}}^{k_{1\cdot N}} \partial_{p_{M}}^l B) (\partial_{q^{1\cdot N}}^{k_{1\cdot N}} \partial_{p_{M}}^l A) \Big)
\\
&\quad- \sum_{\substack{k_{1\cdot N}\\=0}}^0 \prod_{m=1}^N \frac{(-c)^{k_m}}{{k_m}!} \sum_{l=1}^\infty \frac{(-c)^l}{l!} \Big( (\partial_{p_{1\cdot N}}^{k_{1\cdot N}} \partial_{p_{M}}^l A) (\partial_{q^{1\cdot N}}^{k_{1\cdot N}} \partial_{p_{M}}^l B)
\\
&\qquad\qquad\qquad\qquad\qquad\qquad\qquad- (\partial_{p_{1\cdot N}}^{k_{1\cdot N}} \partial_{p_{M}}^l B) (\partial_{q^{1\cdot N}}^{k_{1\cdot N}} \partial_{p_{M}}^l A) \Big)
\\
&\quad- \sum_{\substack{k_{1\cdot N}\\=0*}}^\infty \prod_{m=1}^N \frac{(-c)^{k_m}}{{k_m}!} \sum_{l=1}^\infty \frac{(-c)^l}{l!} \sum_{i,j} (\partial_{p_{1\cdot N}}^{k_{1\cdot N}} \partial_{p_{M}}^l A_i^{1\cdot N} A_i^{M}) (\partial_{q^{1\cdot N}}^{k_{1\cdot N}} \partial_{q^{M}}^l B_j^{1\cdot N} B_j^{M})
\\
&\quad+ \sum_{\substack{k_{1\cdot N}\\=0*}}^\infty \prod_{m=1}^N \frac{(-c)^{k_m}}{{k_m}!} \sum_{l=1}^\infty \frac{(-c)^l}{l!} \sum_{i,j} (\partial_{p_{1\cdot N}}^{k_{1\cdot N}} \partial_{p_{M}}^l B_j^{1\cdot N} B_j^{M}) (\partial_{q^{1\cdot N}}^{k_{1\cdot N}} \partial_{q^{M}}^l A_i^{1\cdot N} A_i^{M})
\\
&= -\sum_{\substack{k_{1\cdot N}\\=0*}}^\infty \sum_{k_M=0}^0 \prod_{m=1}^N \frac{(-c)^{k_m}}{{k_m}!} \frac{(-c)^{k_M}}{{k_M}!} \left( (\partial_{p_{1\cdot M}}^{k_{1\cdot M}} A) (\partial_{q^{1\cdot M}}^{k_{1\cdot M}} B) - (\partial_{p_{1\cdot M}}^{k_{1\cdot M}} B) (\partial_{q^{1\cdot M}}^{k_{1\cdot M}} A) \right)
\\
&\quad- \sum_{\substack{k_{1\cdot N}\\=0}}^0 \sum_{k_M=1}^\infty \prod_{m=1}^N \frac{(-c)^{k_m}}{{k_m}!} \frac{(-c)^{k_M}}{{k_M}!} \left( (\partial_{p_{1\cdot M}}^{k_{1\cdot M}} A) (\partial_{q^{1\cdot M}}^{k_{1\cdot M}} B) - (\partial_{p_{1\cdot M}}^{k_{1\cdot M}} B) (\partial_{q^{1\cdot M}}^{k_{1\cdot M}} A) \right)
\\
&\quad- \sum_{\substack{k_{1\cdot N}\\=0*}}^\infty \sum_{k_M=1}^\infty \prod_{m=1}^N \frac{(-c)^{k_m}}{{k_m}!} \frac{(-c)^{k_M}}{{k_M}!} \left( (\partial_{p_{1\cdot M}}^{k_{1\cdot M}} A) (\partial_{q^{1\cdot M}}^{k_{1\cdot M}} B) - (\partial_{p_{1\cdot M}}^{k_{1\cdot M}} B) (\partial_{q^{1\cdot M}}^{k_{1\cdot M}} A) \right)
\\
&= -\left( \sum_{\substack{k_{1\cdot N}\\=0*}}^\infty \sum_{k_M=0}^0 + \sum_{\substack{k_{1\cdot N}\\=0}}^0 \sum_{k_M=1}^\infty + \sum_{\substack{k_{1\cdot N}\\=0*}}^\infty \sum_{k_M=1}^\infty \right) \prod_{m=1}^M \frac{(-c)^{k_m}}{{k_m}!} 
\\
&\qquad\qquad\cdot \left( (\partial_{p_{1\cdot M}}^{k_{1\cdot M}} A) (\partial_{q^{1\cdot M}}^{k_{1\cdot M}} B) - (\partial_{p_{1\cdot M}}^{k_{1\cdot M}} B) (\partial_{q^{1\cdot M}}^{k_{1\cdot M}} A) \right)
\\
&= -\sum_{\substack{k_{1\cdot M}\\=0*}}^\infty \prod_{m=1}^M \frac{(-c)^{k_m}}{{k_m}!} \left( (\partial_{p_{1\cdot M}}^{k_{1\cdot M}} A) (\partial_{q^{1\cdot M}}^{k_{1\cdot M}} B) - (\partial_{p_{1\cdot M}}^{k_{1\cdot M}} B) (\partial_{q^{1\cdot M}}^{k_{1\cdot M}} A) \right),
\end{align*}
i.e., it is true for $N+1$. Therefore, \eqref{siegelEqwihoutOc2} is true $\forall N\in\mathbb{N}$.

Now, \eqref{siegelEqwihoutOc2} can be rewritten as an explicit power series in $c$, which will be useful to obtain the terms proportional to $c^2,c^3,\ldots$ in \eqref{siegelEq}. Note that
\begin{equation*}
\prod_{m=1}^N \frac{(-c)^{k_m}}{{k_m}!} = (-c)^K\prod_{m=1}^N \frac{1}{{k_m}!} = (-c)^K \frac{1}{{k_1}!\cdots {k_N}!},
\end{equation*}
where $K:=k_1+\cdots+k_N$,\footnote{$K$ is used instead of $k_1+\cdots+k_N$ to fit some large expressions in a page width.} and
\begin{equation*}
\sum_{\substack{k_{1\cdot N}\\=0*}}^\infty \equiv \sum_{k=1}^\infty \sum_{\substack{k_1,\ldots,k_N=0\\\text{such that }K=k}}^k,
\end{equation*}
so
\begin{equation*}
\sum_{\substack{k_{1\cdot N}\\=0*}}^\infty \prod_{m=1}^N \frac{(-c)^{k_m}}{{k_m}!} = \sum_{k=1}^\infty \frac{(-c)^k}{k!} \sum_{\substack{k_1,\ldots,k_N=0\\k_1+\cdots+k_N=k}}^k \frac{(k_1+\cdots+k_N)!}{{k_1}!\cdots {k_N}!}.
\end{equation*}
Therefore,
\begin{equation}\label{sumk1kN}
[A,B] = -\sum_{k=1}^\infty \frac{(-c)^k}{k!} \sum_{\substack{k_{1\cdot N}=0\\K=k}}^k \frac{K!}{{k_1}!\cdots {k_N}!} \left( (\partial_{p_{1\cdot N}}^{k_{1\cdot N}} A) (\partial_{q^{1\cdot N}}^{k_{1\cdot N}} B) - (\partial_{p_{1\cdot N}}^{k_{1\cdot N}} B) (\partial_{q^{1\cdot N}}^{k_{1\cdot N}} A) \right).
\end{equation}
There are $(k+1)^N$ different combinations $(k_1,\ldots,k_N)$ when $k_m$ range from 0 to $k$ in last sum, but only $\frac{(N-1+k)!}{(N-1)!k!}$ of them obey the restriction $k_1+\cdots+k_N=k$. The multinomial coefficient $\frac{(k_1+\cdots+k_N)!}{{k_1}!\cdots {k_N}!}$ (a) counts the number of ways that $k_1+\cdots+k_N$ different elements can be placed in $N$ boxes, when there are $k_m$ elements in box number $m$ and the order of the elements in every box is irrelevant; and (b) obeys
\begin{equation*}
\sum_{\substack{k_1,\ldots,k_N=0\\k_1+\cdots+k_N=k}}^k \frac{(k_1+\cdots+k_N)!}{{k_1}!\cdots {k_N}!}=N^k.
\end{equation*}

Every $k_m$ counts how many partial derivatives of function $A$ (and $B$) are taken respect to variable $p_m$ (and $q^m$) in $(\partial_{p_{1\cdot N}}^{k_{1\cdot N}} A) (\partial_{q^{1\cdot N}}^{k_{1\cdot N}} B) - (\partial_{p_{1\cdot N}}^{k_{1\cdot N}} B) (\partial_{q^{1\cdot N}}^{k_{1\cdot N}} A)$, i.e.,
\begin{equation*}
\partial_{p_{1\cdot N}}^{k_{1\cdot N}} A = \frac{\partial^{k_1}}{\partial {p_1}^{k_1}}\cdots\frac{\partial^{k_N}}{\partial {p_N}^{k_N}}A = \frac{\partial^{k_1+\cdots+k_N} A}{\partial {p_1}^{k_1}\cdots\partial {p_N}^{k_N}},
\end{equation*}
and these partial derivatives can be written in many ways. For example, for $N=4$ and $k=3$, there are $\frac{(N-1+k)!}{(N-1)!k!}=\frac{6!}{3!3!}=20$ different combinations $(k_1,\ldots,k_N)$ that obey $k_1+\cdots+k_N=k$: $(3,0,0,0)$, $(0,3,0,0)$, $(0,0,3,0)$, $(0,0,0,3)$ (these 4 with $\frac{K!}{{k_1}!\cdots {k_N}!}=\frac{3!}{3!}=1$); $(2,1,0,0)$, $(2,0,1,0)$, $(2,0,0,1)$, $(1,2,0,0)$, $(0,2,1,0)$, $(0,2,0,1)$, $(1,0,2,0)$, $(0,1,2,0)$, $(0,0,2,1)$, $(1,0,0,2)$, $(0,1,0,2)$, $(0,0,1,2)$ (these 12 with $\frac{K!}{{k_1}!\cdots {k_N}!}=\frac{3!}{2!}=3$); and $(1,1,1,0)$, $(1,1,0,1)$, $(1,0,1,1)$, $(0,1,1,1)$ (these 4 with $\frac{K!}{{k_1}!\cdots {k_N}!}=\frac{3!}{1!}=6$). Each of these $\frac{(N-1+k)!}{(N-1)!k!}$ combinations corresponds to one addend in the last sum of \eqref{sumk1kN} that is multiplied by a multinomial coefficient which counts the number of ways that partial derivatives can be arranged by using the symmetry property of partial derivatives of ``well-behaved'' functions ($\frac{\partial^2\!f(x,y)}{\partial x\partial y} = \frac{\partial^2\!f(x,y)}{\partial y\partial x}$):\footnote{Since $A=\sum_i A_i^1\cdots A_i^N$, then $\frac{\partial A}{\partial p_m} = \sum_i A_i^1\cdots\frac{\partial A_i^m}{\partial p_m}\cdots A_i^N$. For $m\neq n$ (e.g., $m<n$), $\frac{\partial}{\partial p_n}\frac{\partial A}{\partial p_m} = \sum_i A_i^1\cdots\frac{\partial A_i^m}{\partial p_m}\cdots\frac{\partial A_i^n}{\partial p_n}\cdots A_i^N = \frac{\partial}{\partial p_m}\frac{\partial A}{\partial p_n}$.}
\begin{enumerate}
\item For each of the first 4 combinations, e.g., for $(k_1,\ldots,k_N)=(0,3,0,0)$, the term $(\partial_{p_{1\cdot N}}^{k_{1\cdot N}} A) (\partial_{q^{1\cdot N}}^{k_{1\cdot N}} B) - (\partial_{p_{1\cdot N}}^{k_{1\cdot N}} B) (\partial_{q^{1\cdot N}}^{k_{1\cdot N}} A)$ is multiplied by $\frac{K!}{{k_1}!\cdots {k_N}!}=1$, which is the number of ways that $\partial_{p_{1234}}^{0300} A$ can be written (and $\partial_{q^{1234}}^{0300} B$, $\partial_{p_{1234}}^{0300} B$, $\partial_{q^{1234}}^{0300} A$): $\frac{\partial^3\!A}{\partial {p_2} \partial {p_2} \partial {p_2}}$.
\item For each of the next 12 combinations, e.g., for $(k_1,\ldots,k_N)=(1,0,2,0)$, coefficient $\frac{K!}{{k_1}!\cdots {k_N}!}$ equals 3, which is the number of ways that $\partial_{p_{1234}}^{1020} A$ can be written (and $\partial_{q^{1234}}^{1020} B$, etc.): $\frac{\partial^3\!A}{\partial {p_1} \partial {p_3} \partial {p_3}}$, $\frac{\partial^3\!A}{\partial {p_3} \partial {p_1} \partial {p_3}}$, and $\frac{\partial^3\!A}{\partial {p_3} \partial {p_3} \partial {p_1}}$.
\item For each of the last 4 combinations, e.g., for $(k_1,\ldots,k_N)=(1,1,0,1)$, $\partial_{p_{1234}}^{1101} A$ (and $\partial_{q^{1234}}^{1101} B$, etc.) can be written in $\frac{K!}{{k_1}!\cdots {k_N}!}=6$ different ways: $\frac{\partial^3\!A}{\partial {p_1} \partial {p_2} \partial {p_4}}$, $\frac{\partial^3\!A}{\partial {p_1} \partial {p_4} \partial {p_2}}$, $\frac{\partial^3\!A}{\partial {p_2} \partial {p_1} \partial {p_4}}$, $\frac{\partial^3\!A}{\partial {p_2} \partial {p_4} \partial {p_1}}$, $\frac{\partial^3\!A}{\partial {p_4} \partial {p_1} \partial {p_2}}$, and $\frac{\partial^3\!A}{\partial {p_4} \partial {p_2} \partial {p_1}}$.
\end{enumerate}
This occurs in general for every $N$, $k$ and $(k_1,\ldots,k_N)$ such that $k_1+\cdots+k_N=k$. Therefore, let $n_1,\ldots,n_k$ be $k$ indices that run from 1 to $N$ (thus, there are $N^k$ combinations). Then,
\begin{align*}
\sum_{\substack{k_1,\ldots,k_N=0\\k_1+\cdots+k_N=k}}^{k} & \frac{(k_1+\cdots+k_N)!}{{k_1}!\cdots {k_N}!} \left( (\partial_{p_{1\cdot N}}^{k_{1\cdot N}} A) (\partial_{q^{1\cdot N}}^{k_{1\cdot N}} B) - (\partial_{p_{1\cdot N}}^{k_{1\cdot N}} B) (\partial_{q^{1\cdot N}}^{k_{1\cdot N}} A) \right) \\
= \sum_{n_1,\ldots,n_k=1}^N & \left( \frac{\partial^k\!A}{\partial{p_{n_1}}\cdots\partial{p_{n_k}}} \frac{\partial^k\!B}{\partial{q^{n_1}}\cdots\partial{q^{n_k}}} - \frac{\partial^k\!B}{\partial{p_{n_1}}\cdots\partial{p_{n_k}}} \frac{\partial^k\!A}{\partial{q^{n_1}}\cdots\partial{q^{n_k}}} \right)
\\
\equiv \sum_{n_1,\ldots,n_k=1}^N & \left( \frac{\partial^k\!A}{\partial{p_{n_{1\cdot k}}}}  \frac{\partial^k\!B}{\partial{q^{n_{1\cdot k}}}} - \frac{\partial^k\!B}{\partial{p_{n_{1\cdot k}}}} \frac{\partial^k\!A}{\partial{q^{n_{1\cdot k}}}} \right),
\end{align*}
so
\begin{equation}\label{sumn1nk} 
[A,B] = -\sum_{k=1}^\infty \frac{(-c)^k}{k!} \sum_{n_1,\ldots,n_k=1}^N \left( \frac{\partial^k\!A}{\partial{p_{n_{1\cdot k}}}}  \frac{\partial^k\!B}{\partial{q^{n_{1\cdot k}}}} - \frac{\partial^k\!B}{\partial{p_{n_{1\cdot k}}}} \frac{\partial^k\!A}{\partial{q^{n_{1\cdot k}}}} \right).
\end{equation}

Equation \ref{siegelEqwihoutOc2} is as valid as equations \ref{sumk1kN} or \ref{sumn1nk}, but in \eqref{sumk1kN} and \eqref{sumn1nk} the power series in $c$ (or $\hbar$) is explicit. Also, \eqref{sumk1kN} has less terms in the last sum (concretely, $\frac{(N-1+k)!}{(N-1)!k!}=20$ terms in $(k+1)^N=256$ combinations for $N=4,k=3$) than \eqref{sumn1nk} (which has $N^k=64$ terms for $N=4,k=3$), so less partial derivatives are evaluated in \eqref{sumk1kN} than in \eqref{sumn1nk} if one finds the $\frac{(N-1+k)!}{(N-1)!k!}$ terms. This can be done manually for small $k$'s:
\begin{enumerate}
\item For $k=1$ there are $N$ terms with coefficient 1: they can be computed with an index $m$ running from 1 to $N$, as in \eqref{siegelEq}.
\item For $k=2$ there are $N(N+1)/2$ terms: $N$ with coefficient 1 (for unmixed derivatives) computed with $m\in\{1,\ldots,N\}$, and $N(N-1)/2$ with coefficient 2 (for mixed derivatives) computed with $m\in\{1,\ldots,N-1\},n\in\{m+1,\ldots,N\}$:
\begin{align*}
&\sum_{m=1}^N \left( \frac{\partial^2\!A}{\partial{p_m}^2} \frac{\partial^2\!B}{\partial{q^m}^2} - \frac{\partial^2\!B}{\partial{p_m}^2} \frac{\partial^2\!A}{\partial{q^m}^2} \right)
\\
+ 2 & \sum_{m=1}^{N-1}\sum_{n=m+1}^{N} \left( \frac{\partial^2\!A}{\partial{p_m}\partial{p_n}} \frac{\partial^2\!B}{\partial{q^m}\partial{q^n}} - \frac{\partial^2\!B}{\partial{p_m}\partial{p_n}} \frac{\partial^2\!A}{\partial{q^m}\partial{q^n}} \right).
\end{align*}
Note that $N+N(N-1)/2=N(N+1)/2$, and $1N+2N(N-1)/2=N^2$.
\item For $k=3$ there are $N(N+1)(N+2)/6$ terms: $N$ with coefficient 1 (for unmixed derivatives) computed as always, $N(N-1)$ with coefficient 3 (for simple derivatives of one variable and double derivatives of another) computed with $m\in\{1,\ldots,N\},n\in\{1,\ldots,N\}\backslash\{m\}$, and $N(N-1)(N-2)/6$ with coefficient 6 (for simple derivatives of three different variables) computed with $m\in\{1,\ldots,N-2\},n\in\{m+1,\ldots,N-1\},o\in\{n+1,\ldots,N\}$:
\begin{align*}
&\sum_{m=1}^N \left( \frac{\partial^3\!A}{\partial{p_m}^3} \frac{\partial^3\!B}{\partial{q^m}^3} - \frac{\partial^3\!B}{\partial{p_m}^3} \frac{\partial^3\!A}{\partial{q^m}^3} \right)
\\
+ 3 & \sum_{m=1}^{N}\sum_{\substack{n=1\\(n\neq m)}}^N \left( \frac{\partial^3\!A}{\partial{p_m}\partial{p_n}^2} \frac{\partial^3\!B}{\partial{q^m}\partial{q^n}^2} - \frac{\partial^3\!B}{\partial{p_m}\partial{p_n}^2} \frac{\partial^3\!A}{\partial{q^m}\partial{q^n}^2} \right)
\\
+ 6 & \sum_{m=1}^{N-2}\sum_{n=m+1}^{N-1}\sum_{o=n+1}^{N} \left( \frac{\partial^3\!A}{\partial{p_m}\partial{p_n}\partial{p_o}} \frac{\partial^3\!B}{\partial{q^m}\partial{q^n}\partial{q^o}} - \frac{\partial^3\!B}{\partial{p_m}\partial{p_n}\partial{p_o}} \frac{\partial^3\!A}{\partial{q^m}\partial{q^n}\partial{q^o}} \right).
\end{align*}
Note that $N+N(N-1)+N(N-1)(N-2)/6=N(N+1)(N+2)/6$, and $1N+3N(N-1)+6N(N-1)(N-2)/6=N^3$.
\end{enumerate}
For these $k$'s (and $c=-i\hbar$), this extension of \eqref{siegelEq} can be obtained with  \eqref{sumk1kN}:
\begin{align*}
[A,B] &= -i\hbar\sum_{m=1}^N\left(\frac{\partial A}{\partial p_m}\frac{\partial B}{\partial q^m}-\frac{\partial B}{\partial p_m}\frac{\partial A}{\partial q^m}\right)
\\
&\quad + \frac{\hbar^2}{2} \left( \sum_{\substack{m=1\\(n=m)}}^{N} + 2 \sum_{m=1}^{N-1} \sum_{n=m+1}^{N} \right)
\left( \frac{\partial^2\!A}{\partial{p_m}\partial{p_n}} \frac{\partial^2\!B}{\partial{q^m}\partial{q^n}} - \frac{\partial^2\!B}{\partial{p_m}\partial{p_n}} \frac{\partial^2\!A}{\partial{q^m}\partial{q^n}} \right)
\\
&\quad + \frac{i\hbar^3}{6} \left( \sum_{\substack{m=1\\(n,o=m)}}^{N} + 3 \sum_{m=1}^{N} \sum_{\substack{n=1\\(n\neq m)\\(o=n)}}^{N} + 6 \sum_{m=1}^{N-2}\sum_{n=m+1}^{N-1}\sum_{o=n+1}^{N}\right) 
\\
&\qquad\qquad \left( \frac{\partial^3\!A}{\partial{p_m}\partial{p_n}\partial{p_o}} \frac{\partial^3\!B}{\partial{q^m}\partial{q^n}\partial{q^o}} - \frac{\partial^3\!B}{\partial{p_m}\partial{p_n}\partial{p_o}} \frac{\partial^3\!A}{\partial{q^m}\partial{q^n}\partial{q^o}} \right) + \mathcal{O}(\hbar^4).
\end{align*}

With the goal of this work accomplished, please note the structure of \eqref{fundamentalEq}: it is an $N$-dimensional infinite sum that equals zero, whose $(0,\ldots,0)$th term is the commutator of interest. In fact, all equations regarding commutators share this structure, with $[A,B]$ being the 0th term of a sum equal to zero.

\section{Extension to inverses}

In this section, \eqref{sumk1kN} will be extended to include functions $A$ and $B$ that depend on negative powers of $p_m$ and $q^m$, provided that $\mathcal{A}$ has some extra properties not previously needed. As above, suppose that $\exists p,q\in\mathcal{A}$ such that $[p,q]=cI$ for some $c\in\mathcal{F}$. Now, if $\exists A\in\mathcal{A}$ such that $Ap=pA=I$ ($p$ has a multiplicative inverse), then it can denoted as $p^{-1}$ or $I/p$, and it can be proved that $p^{-i}:=\underbrace{p^{-1} \cdots p^{-1}}_{i\text{ times}}\in\mathcal{A}\;\forall i\in\mathbb{N}$ (because $\mathcal{A}$ is closed under multiplication), $p^{x+y} = p^x p^y \;\forall x,y\in\mathbb{Z}$, and $p^{-x}$ is the inverse of $p^x\;\forall x\in\mathbb{Z}$. Also, if $\exists A\in\mathcal{A}$ such that $Aq=qA=I$, then similar statements can be made for $q^{x+y}$ and $q^{-x}$ ($x,y\in\mathbb{Z}$). In this section it is supposed that $p^{-1},q^{-1},{(p_m)}^{-1},{(q^m)}^{-1}\in\mathcal{A}$.

First step is to compute $[p^{-1},q]=(I/p)q-q/p$ and $[p,q^{-1}]=p/q-(I/q)p$:
\begin{align*}
\text{As }p[p^{-1},q] &= q-pq/p=q-(qp+[p,q])/p=q-q-cI/p=-cp^{-1},
\\
\text{then }[p^{-1},q] &= p^{-1}(-cp^{-1})=-cp^{-2}=:c\partial_p p^{-1}.
\\
\text{Also, }q[p,q^{-1}] &= qp/q-p=(pq-[p,q])/q-p=p-cq^{-1}-p=-cq^{-1},
\\
\text{so }[p,q^{-1}] &= q^{-1}(-cq^{-1})=-cq^{-2}=:c\partial_q q^{-1}.
\end{align*}
Therefore,
\begin{align*}
\text{if }[p^{-i},q] &= -cip^{-i-1}=:c\partial_p p^{-i}\text{, then}
\\
[p^{-i-1},q] &= [p^{-i},q]p^{-1}+p^{-i}[p^{-1},q] = -cip^{-i-1}p^{-1}+p^{-i}(-cp^{-2})
\\
&=-c(i+1)p^{-(i+1)-1}=:c\partial_p p^{-(i+1)} \quad \forall i\in\mathbb{N};
\\
\text{if }[p,q^{-j}] &= -cjq^{-j-1}=:c\partial_q q^{-j}\text{, then}
\\
[p,q^{-j-1}] &= [p,q^{-j}]q^{-1}+q^{-j}[p,q^{-1}] = -cjq^{-j-1}q^{-1}+q^{-j}(-cq^{-2})
\\
&=-c(j+1)q^{-(j+1)-1}=:c\partial_q q^{-(j+1)} \quad \forall j\in\mathbb{N}.
\end{align*}
Since it was proved that $[p^i,q]=c\partial_p p^i,[p,q^j]=c\partial_q q^j\;\forall i,j\in\mathbb{N}$, then
\begin{equation*}
[p^x,q]=cxp^{x-1}=c\partial_p p^x\text{ and }[p,q^y]=cyq^{y-1}=c\partial_q q^y\;\forall x,y\in\mathbb{Z}.
\end{equation*}

Now, next goal is to obtain the equivalent to \eqref{ABwithoutOc2} when $A^p$ (and $B^q$) is a linear combination of integer powers of $p$.
\begin{align*}
[p^2,q^y] &= [p,q^y]p+p[p,q^y] = cyq^{y-1}p+pcyq^{y-1}
\\
&= cy(pq^{y-1}-[p,q^{y-1}])+cypq^{y-1} = 2cypq^{y-1}-c^2y(y-1)q^{y-2}
\\
&= c(\partial_p p^2)(\partial_q q^y) -\frac{c^2}{2}(\partial_p^2 p^2)(\partial_q^2 q^y),
\end{align*}
which agrees with \eqref{ABwithoutOc2} for $A^p=p^2$ and $B^q=q^y$. Therefore, the induction hypothesis on $x$ will be
\begin{equation}\label{pxqy}
[p^x,q^y] = -\sum_{k=1}^\infty \frac{(-c)^k}{k!}(\partial_p^k p^x)(\partial_q^k q^y).
\end{equation}
Note that $\partial_p^k p^x := x(x-1)\cdots(x-k+1) p^{x-k}$ and $\partial_q^k q^y := y(y-1)\cdots(y-k+1)q^{y-k}$, so if \eqref{pxqy} is true\footnote{This infinite sum has finitely many non-zero terms for a non-negative integer $x$ (or $y$, or both). If both $x$ and $y$ are negative integers, the series should converge to $p^xq^y-q^yp^x\in\mathcal{A}$ whenever $\exists p^{-1},q^{-1}\in\mathcal{A}$.} for some $x,y\in\mathbb{Z}$, then
\begin{align}\label{dipxidjqyj}
[\partial_p^i p^{x+i},\partial_q^j q^{y+j}] &= [(x+i)\cdots(x+1)p^x,(y+j)\cdots(y+1)q^y]\notag
\\
&= -(x+i)\cdots(x+1)(y+j)\cdots(y+1)\sum_{k=1}^\infty \frac{(-c)^k}{k!}(\partial_p^k p^x)(\partial_q^k q^y)\notag
\\
&= -\sum_{k=1}^\infty \frac{(-c)^k}{k!}\left((x+i)\cdots(x+1)\partial_p^k p^x\right) \left((y+j)\cdots(y+1)\partial_q^k q^y\right)\notag
\\
&= -\sum_{k=1}^\infty \frac{(-c)^k}{k!}(\partial_p^{k+i} p^{x+i})(\partial_q^{k+j} q^{y+j})
\end{align}
is also true for the same $x,y$ and $\forall i,j\in\mathbb{N}$, so whenever we had proved \eqref{pxqy} for a given pair $x,y$, \eqref{dipxidjqyj} could be equally applied for the same $x,y$. Conversely, \eqref{dipxidjqyj} reduces to \eqref{pxqy} for $i=j=0$. Note that \eqref{pxqy} already holds for $x=0,1,2$ and any $y$, so \eqref{dipxidjqyj} also does.

Before starting to prove \eqref{pxqy}, a result that will be used must be checked first: Leibnitz formula for the derivatives of a product might not hold for ``derivatives'' of ``functions'' on one-element sets. Therefore, let $A$ be any element in $\mathcal{A}$, $x,y$ any two elements in $\mathbb{Z}$, and $i,j$ any two elements in $\mathbb{N}$; then
\begin{align*}
\partial_A (A^xA^y) &= \partial_A A^{x+y} := (x+y)A^{x+y-1}=xA^{x-1}A^y+A^xyA^{y-1}
\\
&= (\partial_A A^x)A^y+A^x(\partial_A A^y);
\\
\partial_A \big( (\partial_A^i A^x)(\partial_A^j A^y) \big) &= x\cdots(x-i+1)y\cdots(y-j+1)\partial_A (A^{x-i}A^{y-i})
\\
&= x\cdots(x-i+1)y\cdots(y-j+1)(\partial_A A^{x-i})A^{y-j}
\\
&\quad+ x\cdots(x-i+1)y\cdots(y-j+1)A^{x-i}(\partial_A A^{y-j})
\\
&= (\partial_A \partial_A^i A^x)(\partial_A^j A^y) + (\partial_A^i A^x)(\partial_A \partial_A^j A^y)
\\
&= (\partial_A^{i+1} A^x)(\partial_A^j A^y) + (\partial_A^i A^x)(\partial_A^{j+1} A^y);
\\
\text{if }\partial_A^k (A^xA^y) &= \sum_{l=0}^k \binom{k}{l} (\partial_A^{k-l} A^x)(\partial_A^l A^y)\text{, then}
\\
\partial_A^{k+1} (A^xA^y) &= \partial_A \sum_{l=0}^k \binom{k}{l} (\partial_A^{k-l} A^x)(\partial_A^l A^y)
\\
&= \sum_{l=0}^k \binom{k}{l} \left( (\partial_A^{k-l+1} A^x)(\partial_A^l A^y) + (\partial_A^{k-l} A^x)(\partial_A^{l+1} A^y) \right)
\\
&= \sum_{l=0}^k \binom{k}{l} (\partial_A^{k-l+1} A^x)(\partial_A^l A^y) + \sum_{l=1}^{k+1} \binom{k}{l-1} (\partial_A^{k-l+1} A^x)(\partial_A^{l} A^y)
\\
&= (\partial_A^{k+1} A^x)(A^y) + \sum_{l=1}^k \binom{k+1}{l} (\partial_A^{k-l+1} A^x)(\partial_A^l A^y) + (A^x)(\partial_A^{k+1} A^y)
\\
&= \sum_{l=0}^{k+1} \binom{k+1}{l} (\partial_A^{k+1-l} A^x)(\partial_A^l A^y),
\end{align*}
so Leibnitz formula holds for $\partial_A^k(A^xA^y)\;\forall k\in\mathbb{N}$. Note that Leibnitz formula for $\partial_A^k\left(f(A)g(A)\right)$ can be proven easily with this result when $f(A)=\sum_x f_x A^x$, $g(A)=\sum_y g_y A^y$, $f_x,g_y\in\mathcal{F}$, $x,y\in\mathbb{Z}$.

Now we are ready to prove \eqref{pxqy}. First, we begin with $x=-1$:
\begin{align*}
[p^{-1},q^{y}] &= p^{-1}q^{y}-q^{y}p^{-1}\text{, so}
\\
p[p^{-1},q^{y}]p &= q^{y}p-pq^{y} = -[p,q^{y}] = -cyq^{y-1}.
\end{align*}
Thus, now it will be checked that right-hand side of \eqref{pxqy} for $x=-1$ equals $-cyq^{y-1}$ when it is left- and right-multiplied by $p$:
\begin{align*}
\partial_p^k p^{-i} &= (-i)\cdots(-i-k+1)p^{-i-k} = (-1)^{k}\frac{(i+k-1)!}{(i-1)!}p^{-i-k}\text{, so}
\\
\partial_p^k p^{-1} &= (-1)^{k} k! p^{-k-1};
\\
\text{let }A &:= -\sum_{k=1}^\infty \frac{(-c)^k}{k!}(\partial_p^k p^{-1})(\partial_q^k q^{y})\text{, then}
\\
pAp &= -p\sum_{k=1}^\infty \frac{(-c)^k}{k!} (-1)^k k! p^{-k-1} y(y-1)\cdots(y-k+1) q^{y-k} p
\\
&= -\sum_{k=1}^\infty c^k y(y-1)\cdots(y-k+1)p^{-k} \left(pq^{y-k} - [p,q^{y-k}]\right)
\\
&= -\sum_{k=1}^\infty c^k y(y-1)\cdots(y-k+1) \left(p^{-k+1}q^{y-k} - c(y-k)p^{-k}q^{y-k-1}\right)
\\
&= -c^1yp^0q^{y-1} -\underbrace{\sum_{k=2}^\infty c^k y(y-1)\cdots(y-k+1) p^{-k+1}q^{y-k}}_{\text{this term equals\ldots}}
\\
&\quad+ \underbrace{\sum_{k=1}^\infty c^{k+1} y(y-1)\cdots(y-k) p^{-k}q^{y-k-1}}_{\text{\ldots this term}}
\\
&= -cyq^{y-1}\text{; then,}
\\
A &= -p^{-1}cyq^{y-1}p^{-1} = p^{-1}p[p^{-1},q^{y}]pp^{-1} = [p^{-1},q^{y}].
\end{align*}
Therefore, \eqref{pxqy} holds for $x=-1$ and any $y$. Lastly, suppose \eqref{pxqy} true for $x$ and prove it for $x+z$, where $z=-1,0,1,2$ (the values of $x$ for which \eqref{pxqy} is already proved for any $y$):
\begin{align*}
[p^{x+z},q^y] &= [p^{x},q^y]p^{z}+p^{x}[p^{z},q^y]
\\
&= -\sum_{k=1}^\infty \frac{(-c)^k}{k!}(\partial_p^k p^{x})(\partial_q^k q^{y})p^{z} -p^{x}\sum_{k=1}^\infty \frac{(-c)^k}{k!}(\partial_p^k p^{z})(\partial_q^k q^{y})
\\
&= -\sum_{k=1}^\infty \frac{(-c)^k}{k!} \left( (\partial_p^k p^{x}) \left( p^{z}(\partial_q^k q^{y})-[p^{z},\partial_q^k q^{y}] \right) + p^{x}(\partial_p^k p^{z})(\partial_q^k q^{y}) \right)
\\
&= -\sum_{k=1}^\infty \frac{(-c)^k}{k!} \Big( (\partial_p^k p^{x}) \big( p^{z}(\partial_q^k q^{y})+\sum_{l=1}^\infty\frac{(-c)^l}{l!}(\partial_p^l p^{z})(\partial_q^{l+k} q^{y}) \big)
\\
&\qquad\qquad+ p^{x}(\partial_p^k p^{z})(\partial_q^k q^{y}) \Big)
\\
&= -\sum_{k=1}^\infty\sum_{l=1}^\infty \frac{(-c)^k}{k!} \frac{(-c)^l}{l!} (\partial_p^k p^{x})(\partial_p^l p^{z})(\partial_q^{k+l} q^{y})
\\
&\quad-\sum_{k=1}^\infty \frac{(-c)^k}{k!} \left( (\partial_p^k p^{x}) p^{z}(\partial_q^k q^{y}) + p^{x}(\partial_p^k p^{z})(\partial_q^k q^{y}) \right)
\\
&= -\sum_{\substack{k=2\\(k+l\mapsto k)}}^\infty\sum_{l=1}^{k-1} \frac{(-c)^{k}}{(k-l)!l!} (\partial_p^{k-l} p^{x})(\partial_p^l p^{z})(\partial_q^{k} q^{y})
\\
&\quad-\sum_{k=1}^\infty \frac{(-c)^k}{k!} \left( (\partial_p^k p^{x}) p^{z} + p^{x}(\partial_p^k p^{z}) \right) (\partial_q^k q^{y})
\\
&= -\sum_{k=1}^\infty \frac{(-c)^{k}}{k!} \underbrace{\sum_{l=1}^{k-1} \frac{k!}{(k-l)!l!} (\partial_p^{k-l} p^{x})(\partial_p^l p^{z})}_{=0\text{ for }k=1} (\partial_q^{k} q^{y})
\\
&\quad-\sum_{k=1}^\infty \frac{(-c)^k}{k!} \Big( \underbrace{(\partial_p^k p^{x})(\partial_p^0 p^{z})}_{l=0} + \underbrace{(\partial_p^0 p^{x})(\partial_p^k p^{z})}_{l=k} \Big) (\partial_q^k q^{y})
\\
&= -\sum_{k=1}^\infty \frac{(-c)^{k}}{k!} \sum_{l=0}^{k} \binom{k}{l} (\partial_p^{k-l} p^{x})(\partial_p^l p^{z}) (\partial_q^{k} q^{y})
\\
&= -\sum_{k=1}^\infty \frac{(-c)^{k}}{k!} (\partial_p^{k} p^{x} p^{z}) (\partial_q^{k} q^{y}) = -\sum_{k=1}^\infty \frac{(-c)^{k}}{k!} (\partial_p^{k} p^{x+z}) (\partial_q^{k} q^{y}),
\end{align*}
i.e., it is true for $x+z$ where $z$ can be $1$ or $-1$. Therefore, \eqref{pxqy} and \eqref{dipxidjqyj} hold for any integers $x$ and $y$. Also, trivially $x+i,y+j\in\mathbb{Z}$ $\forall x,y\in\mathbb{Z}$ $\forall i,j\in\mathbb{N}$, thus by replacing $x+i,y+j$ by $x,y$ in \eqref{dipxidjqyj} one gets
\begin{equation*}\label{dipxdjqy}
[\partial_p^i p^{x},\partial_q^j q^{y}] = -\sum_{k=1}^\infty \frac{(-c)^k}{k!}(\partial_p^{k+i} p^{x})(\partial_q^{k+j} q^{y})\quad\forall x,y\in\mathbb{Z},\forall i,j\in\mathbb{N}.
\end{equation*}
Now, let $A^r(r):=\sum_{x\in\mathbb{Z}}a_x^r r^x$ and $B^r(r):=\sum_{y\in\mathbb{Z}}b_y^r r^y$ ($r$ stands for $p$ or $q$ at convenience, $a_x,b_y\in\mathcal{F}$), which supersedes the definitions of $A^p(p)$, $A^q(q)$, $B^p(p)$ and $B^q(q)$ at the beginning of section 2. Then,
\begin{align}\label{ApBqIntegerExponents}
[A^p,B^q] &= \sum_{x\in\mathbb{Z}} \sum_{y\in\mathbb{Z}} a_x^p b_y^q [p^x,q^y]
\notag\\
&= -\sum_{x\in\mathbb{Z}} \sum_{y\in\mathbb{Z}} a_x^p b_y^q \sum_{k=1}^\infty \frac{(-c)^k}{k!}(\partial_p^k p^x)(\partial_q^k q^y)
\notag\\
&= -\sum_{k=1}^\infty \frac{(-c)^k}{k!}\left(\partial_p^k \sum_{x\in\mathbb{Z}} a_x^p p^x\right) \left(\partial_q^k \sum_{y\in\mathbb{Z}} b_y^q q^y\right)
\notag\\
&= -\sum_{k=1}^\infty \frac{(-c)^k}{k!}\left(\partial_p^k A^p \right) \left(\partial_q^k B^q \right),
\end{align}
which is the equivalent to \eqref{ABwithoutOc2} when $A^p$ ($B^q$) is a linear combination of integer powers of $p$ ($q$).

Next, let $A(p,q):=\sum_i A_i^p(p)A_i^q(q)$ and $B(p,q):=\sum_j B_j^p(p)B_j^q(q)$. Note that the procedure following \eqref{ABwithoutOc2} does not use the explicit nature of $A_i^p(p)$, $A_i^q(q)$, $B_j^p(p)$ and $B_j^q(q)$, so everything from \eqref{ABwithoutOc2} to the end of section 3 can be obtained from \eqref{ApBqIntegerExponents} for integer powers instead of natural powers.

\bibliographystyle{livrevrel} 
\bibliography{postmaster} 

\begin{thebibliography}{1}

\bibitem{Dirac:1925}
{Dirac}, P.~A.~M., ``{The Fundamental Equations of Quantum Mechanics}'', {\em
  Proceedings of the Royal Society of London Series A}, {\bf 109}(752),
  642--653 (December 1925).
  {\small[\href{http://dx.doi.org/10.1098/rspa.1925.0150}{DOI}]},
  {\small[\href{https://ui.adsabs.harvard.edu/abs/1925RSPSA.109..642D}{ADS}]}.

\bibitem{Siegel:Fields1}
{Siegel}, Warren, {\em {Fields v1}}, (1999).
  {\small[\href{https://ui.adsabs.harvard.edu/abs/1999hep.th...12205S}{ADS}]},
  {\small[\href{http://arxiv.org/abs/hep-th/9912205}{{arXiv:hep-th/9912205
  {\small[hep-th]}}}]}.

\bibitem{Siegel:Fields4}
{Siegel}, Warren, {\em {Fields v4}}, (2021).
  {\small[\href{http://insti.physics.sunysb.edu/~siegel/errata.shtml}{{CNYITP,SUNY:~siegel/errata.shtml}}]}.

\bibitem{TranstrumVanhuele:2005}
{Transtrum}, Mark~K.  and {Van Huele}, Jean-Fran{\c{c}}ois~S., ``{Commutation
  relations for functions of operators}'', {\em Journal of Mathematical
  Physics}, {\bf 46}(6), 063510 (June 2005).
  {\small[\href{http://dx.doi.org/10.1063/1.1924703}{DOI}]},
  {\small[\href{https://ui.adsabs.harvard.edu/abs/2005JMP....46f3510T}{ADS}]}.

\end{thebibliography}

\end{document}